\documentclass[9pt,twocolumn,twoside]{opticajnl}
\journal{opticajournal} 

\setboolean{shortarticle}{true}

\usepackage{lineno}

\title{Tunable 30 GHz laser frequency comb for astronomical spectrograph characterization and calibration}

\author[1,2,*]{Pooja Sekhar}
\author[1,2]{Molly Kate Kreider}
\author[1,2,3]{Connor Fredrick}
\author[4]{Joe P. Ninan}
\author[5]{Chad F Bender}
\author[6]{Ryan Terrien}
\author[7,8]{Suvrath Mahadevan}
\author[1,2,\dag]{Scott A. Diddams}

\affil[1]{Electrical, Computer and Energy Engineering, University of Colorado Boulder, Colorado 80309, USA}
\affil[2]{Department of Physics, University of Colorado Boulder, Colorado 80309, USA}
\affil[3]{National Institute of Standards and Technology, Boulder, Colorado 80305, USA}
\affil[4]{Department of Astronomy and Astrophysics, Tata Institute of Fundamental Research (TIFR),
Mumbai, Maharashtra 400005, India}
\affil[5]{Steward Observatory, University of Arizona, Tucson, Arizona 85721, USA}
\affil[6]{Department of Physics and Astronomy, Carleton College, Northfield, Minnesota 55057, USA}
\affil[7]{Department of Astronomy and Astrophysics, Pennsylvania State University, University Park, PA 16802, USA}
\affil[8]{Center for Exoplanets and Habitable Worlds, Pennsylvania State University, University Park, PA 16802, USA}
\affil[*]{Corresponding author: pooja.sekhar@colorado.edu}

\affil[$\dag$]{email: scott.diddams@colorado.edu}

\begin{abstract}
The search for earth-like exoplanets with the Doppler radial velocity technique is an extremely challenging and multifaceted precision spectroscopy problem. Currently, one of the limiting instrumental factors in reaching the required long-term $10^{-10}$ level of radial velocity precision is the defect-driven sub-pixel quantum efficiency variations in the large-format detector arrays used by precision echelle spectrographs. Tunable frequency comb calibration sources that can fully map the point spread function across a spectrograph’s entire bandwidth are necessary for quantifying and correcting these detector artifacts. In this work, we demonstrate a combination of laser frequency and mode spacing control that allows full and deterministic tunability of a 30 GHz electro-optic comb together with its filter cavity. After supercontinuum generation, this gives access to any optical frequency across 700 - 1300 nm.  Our specific implementation is intended for the comb deployed at the Habitable Zone Planet Finder spectrograph and its near-infrared Hawaii-2RG array, but the techniques apply to all laser frequency combs used for precision astronomical spectrograph calibration and other applications that require broadband tuning. 
\end{abstract}

\setboolean{displaycopyright}{false} 

\begin{document}

\maketitle

\begin{figure*}[htb]
\centering
\includegraphics[width=\linewidth]{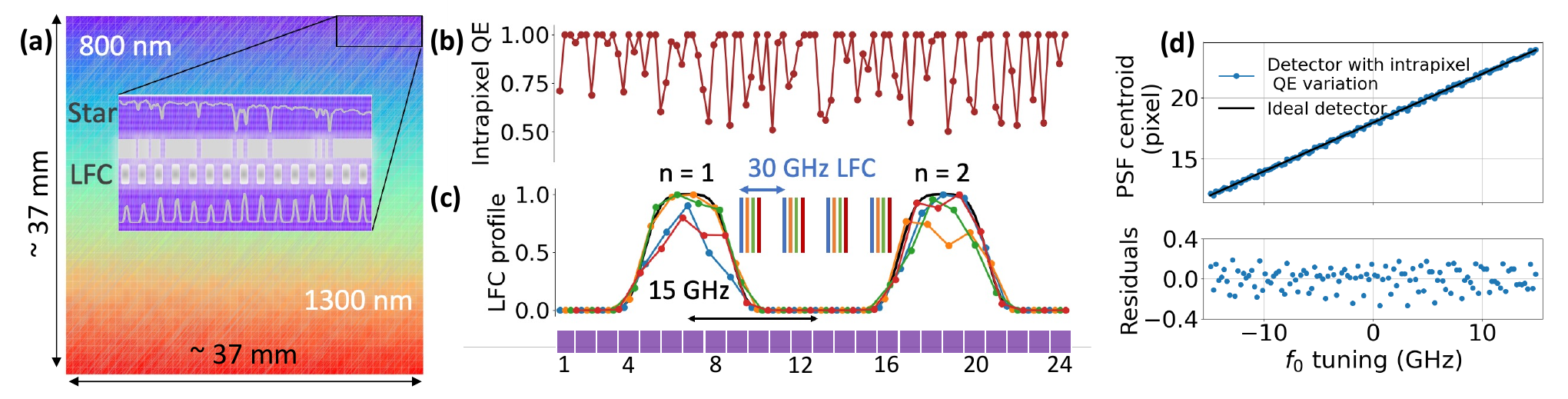}
\caption{The need for a tunable laser frequency comb (LFC) for detector characterization. (a) Sketch of HPF detector array with cross-hatch patterns, with zoom-in showing an example stellar and comb spectra. (b) Simulated variation in intrapixel quantum efficiency (QE). 24 pixels are each sub-divided by $4\times$ and the variation in QE is exaggerated for better visualization. (c) Point spread function (PSF) of two adjacent comb lines on the detector array. The black curve corresponds to the ideal detector with no defects. PSF defined by blue, orange, green and red traces are obtained by tuning the comb frequency in steps of 625 MHz spanning a single pixel. The intrapixel QE is directly mapped onto the PSF with comb tuning. (d) Modeled variation in the centroid of PSF (blue dots) in the presence of cross-hatch patterns as the comb is tuned. The black curve corresponds to the ideal linear variation of the centroid of PSF, and residuals are shown below. If left uncorrected, this can lead to undesirable variations of radial-velocity signals.}
\label{fig:1}
\end{figure*}

\begin{figure*}[hbt!]
\centering
\includegraphics[width=0.92\linewidth]{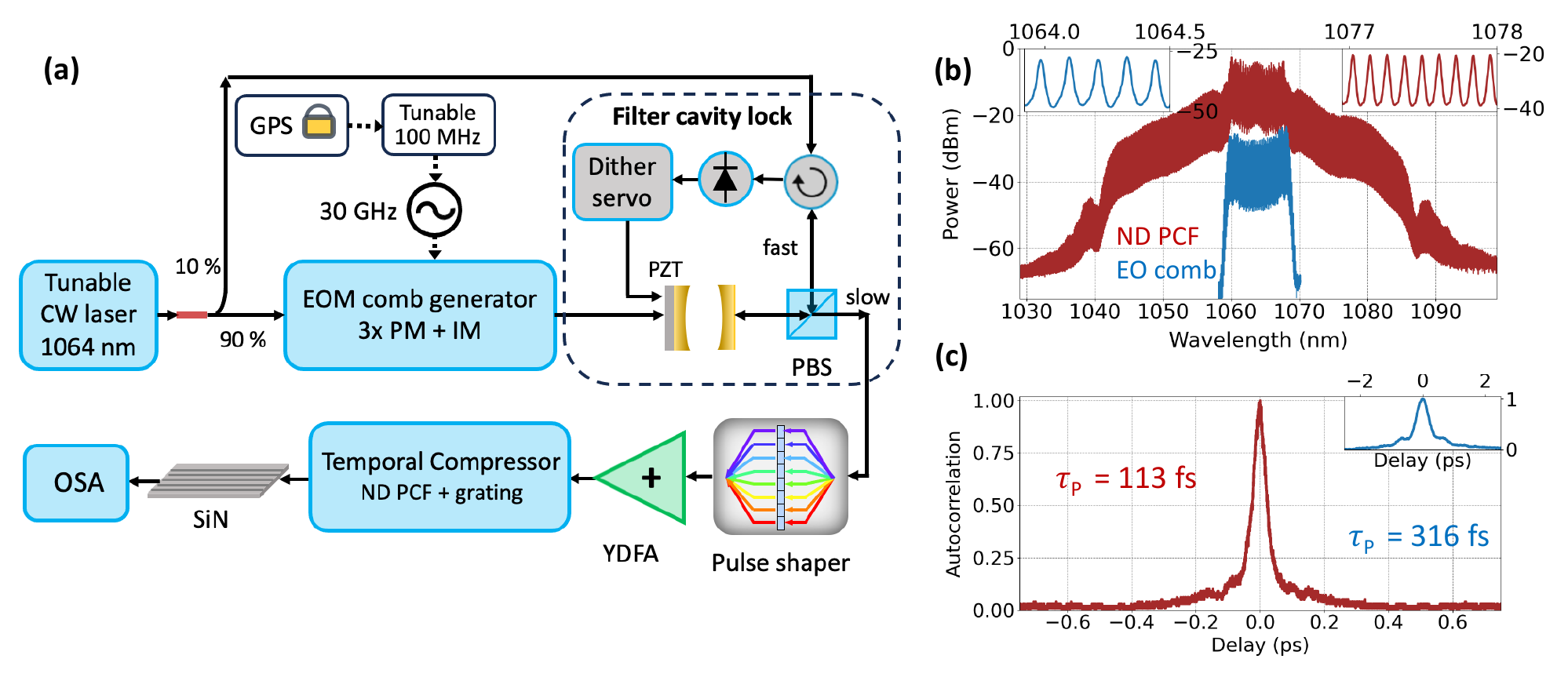}
\caption{(a) Experimental setup for generating a tunable 30 GHz comb spanning 700 - 1350 nm. GPS: global positioning system, YDFA:  ytterbium-doped fiber amplifier, ND PCF: normal dispersion photonic crystal fiber, SiN: silicon nitride. (b) Spectral output of the initial 30 GHz EO comb after the pre-amplifier (blue trace) and temporal compressor consisting of ND PCF and grating (red trace). The traces are vertically offset for clarity. (c) Intensity autocorrelation trace of the 30 GHz comb output (blue) and after the temporal compressor (red). The measured pulse durations are labeled.}
\label{fig:2}
\end{figure*}
Laser frequency combs (LFC) with tens of gigahertz (GHz) mode spacings and bandwidths spanning more than 500 nm have emerged as ideal calibration sources for radial velocity (RV) astronomical spectrographs for exoplanet detection \cite{osterman2007proposed, herr2019astrocombs, steinmetz2008laser}. The rate of exoplanet detection by the RV technique has been steadily increasing, but finding an Earth-mass planet in the habitable zone of a Sun-like star, which requires an RV precision below 10 cm/s, is still very challenging. Given the typical resolution and dispersion of these astronomical spectrographs, this spectral shift amounts to accurately finding line center shifts within a few atoms on the detector lattice. This high level of RV precision is presently limited by hard-to-solve problems like stellar activity and telluric absorption, but the impact of these cannot be fully assessed and ameliorated until the instrumental precision is sufficient to measure and disentangle their impact. Currently, detector inhomogeneities are a major impediment to reaching this level of instrumental precision. Even the best near-infrared HgCdTe detectors and visible silicon charge-coupled devices (CCD) present significant barriers to precision spectroscopy through the brighter-fatter effect, nonlinear response, stitching errors, and intra- and inter-pixel quantum efficiency (QE) variation \cite{lage2017measurements, plazas2017nonlinearity, ninan2019impact}. Thus, detailed characterization of the line profile at all points of the detector is necessary to improve spectral fidelity and enable the detection of Earth-analogs \cite{ninan2019impact,zhao2021measuring, wilken2010high, schmidt2024characterization}. 

As a specific example, the Habitable-Zone Planet Finder (HPF) is an ultrastable fiber-fed near-infrared (NIR: 810 - 1280 nm) spectrograph, which has, along with our 30 GHz LFC calibrator, demonstrated levels of on-sky RV precision of 1.5 m/s \cite{mahadevan2012habitable, mahadevan2014habitable, metcalf2019stellar}. But its NIR Teledyne H2RG detector array suffers from subpixel QE variation due to lattice defects in the HgCdTe layer \cite{shapiro2018intra}. These appear as cross-hatch patterns on the detector as shown in Fig. \ref{fig:1}(a). As a result, the traditional flat correction scheme fails for a non-uniform stellar or LFC source, since the required correction is a function of the intensity distribution within a pixel. Ninan \textit{et. al.} \cite{ninan2019impact} have modeled the impact of these detector artifacts on RV precision (estimated to be 0.4 m/s on HPF) and show that a tunable calibration source can be used to quantify and counteract these wavelength calibration errors. 

Figure \ref{fig:1}(b)-(d) conceptually illustrates the impact of cross-hatch and intrapixel QE variation within a small region of the detector. When the amplitude of any non-uniform spectral source, like a frequency comb, varies within a pixel the point spread function (PSF) at the detector read-out changes in shape following the QE variation within pixels. This results in a change in the centroid of the PSF at different location on the detector (Fig. \ref{fig:1}(d)). This changing centroid of the PSF in turn maps to a wavelength calibration error and subsequently amounts to a large source of RV error, comparable to the required precision levels. To characterize all the pixels similarly, we need to tune the comb frequency by the full comb repetition rate (30 GHz here). 

In this work, we address this challenge by extending the 30 GHz electro-optic (EO) comb to a dynamic metrological tool that can map the point spread function across the entire spectrograph bandwidth. This is achieved by tuning a combination of laser frequency and mode spacing. While demonstrated here with an EO comb, this tandem tuning technique is applicable for both microcombs \cite{del2011octave} and mode-filtered combs \cite{zhao2021measuring, wilken2010high, schmidt2024characterization} that need full free spectral range (FSR) tuning. Variations in the line profile of spectrographs with position and comb line intensities have previously been characterized over a limited range \cite{zhao2021measuring, wilken2010high,ninan2019impact, schmidt2024characterization} or with poor contrast \cite{obrzud2024astrocomb}. However, to the best of our knowledge, this is the first time a tunable >10 GHz comb source spanning nearly an octave has been developed to fully map the entire detector array as a critical step towards improved RV precision. This tandem tuning technique can also be used in direct frequency comb spectroscopy and other applications that require the precise positioning of the comb lines to match atomic or molecular frequencies. \cite{diddams2007molecular, silfies2020widely}.

Fig. \ref{fig:2}(a) shows the schematic of our experimental setup, which provides a comb similar to that deployed at the HPF \cite{metcalf201930}. A tunable free-running continuous-wave (CW) laser at 1064 nm is modulated by a series of three fiber-integrated polarization-maintaining phase modulators (PM) and an intensity modulator (IM). The frequency of the 30 GHz radio-frequency (RF) source can be fine-tuned linearly by adjusting its 100 MHz reference, which is frequency-stabilized to the global positioning system (GPS) via a 10 MHz Rb atomic clock. 
The resultant flat spectrum spanning 2.4 THz after a Yb preamplifier is shown in Fig. \ref{fig:2}(b). The quasi-linear chirp from the PMs and dispersion from the fibers are compensated by a pulse shaper \cite{otsuji199610} to yield a pulse duration of approximately 316 fs (Fig. \ref{fig:2}(c)).  

The EO comb generation is followed by a Fabry-P\'erot (FP) cavity with an FSR that matches the comb repetition rate of 30 GHz. This is necessary to suppress the broadband noise between the comb lines arising from the thermal noise of the high-power RF amplifiers and to maintain the coherence after optical amplification and nonlinear spectral broadening \cite{beha2017electronic, carlson2018ultrafast}. The purpose-built fiber-coupled FP cavity is constructed from highly reflective mirrors (R = 99.5 $\%$) with radii of curvature of 50 cm and 75 cm respectively \cite{ycas2013laser}. The cavity linewidth at full width at half-maximum (FWHM) and finesse are measured to be 64 MHz and 470 respectively. The filter cavity length is locked to the CW laser frequency using a commercial FPGA (Red Pitaya) via the dither technique as shown in Fig. \ref{fig:2}(a) \cite{luda2019compact}. 
\begin{figure}[h!]
\centering
\includegraphics[width=0.99\linewidth]{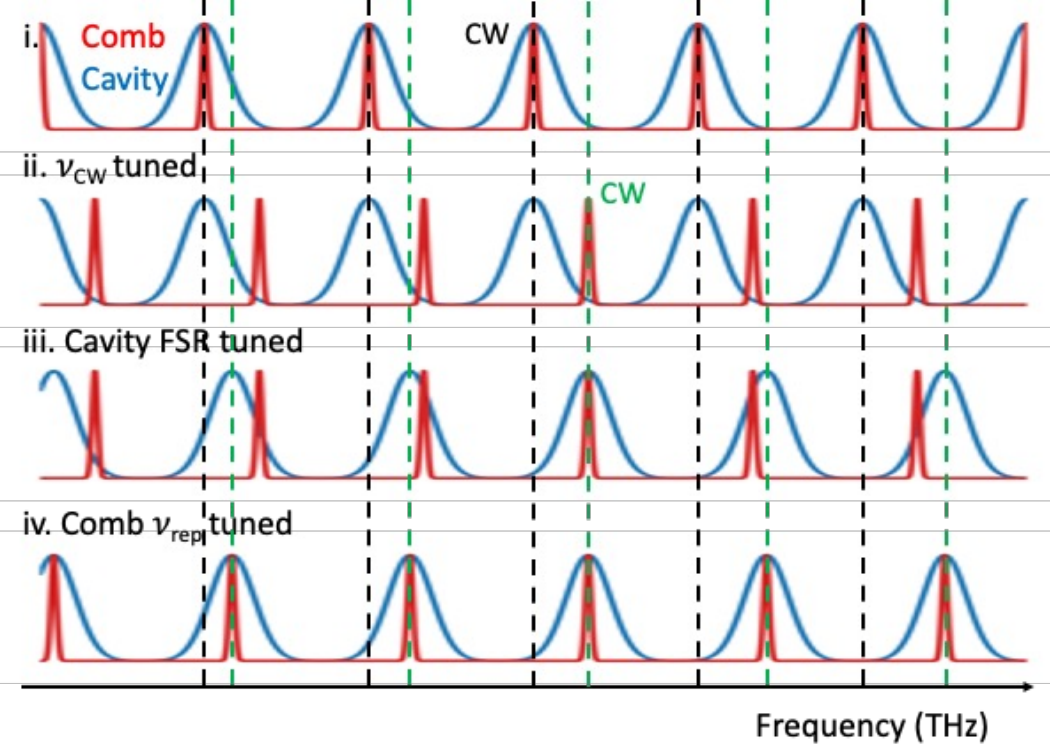}
\caption{Tuning the EO comb. (i) Initial comb and filter cavity. (ii) Tuning the CW laser frequency. (iii) The filter cavity servo adjusts the cavity length to match the CW laser. (iv) The comb repetition rate is accordingly tuned to avoid cavity-comb mode walk-off. The black and green dashed lines indicate the initial and final comb mode positions.}
\label{fig:3}
\end{figure}
\begin{figure}[h!]
\centering
\includegraphics[width=0.8\linewidth]{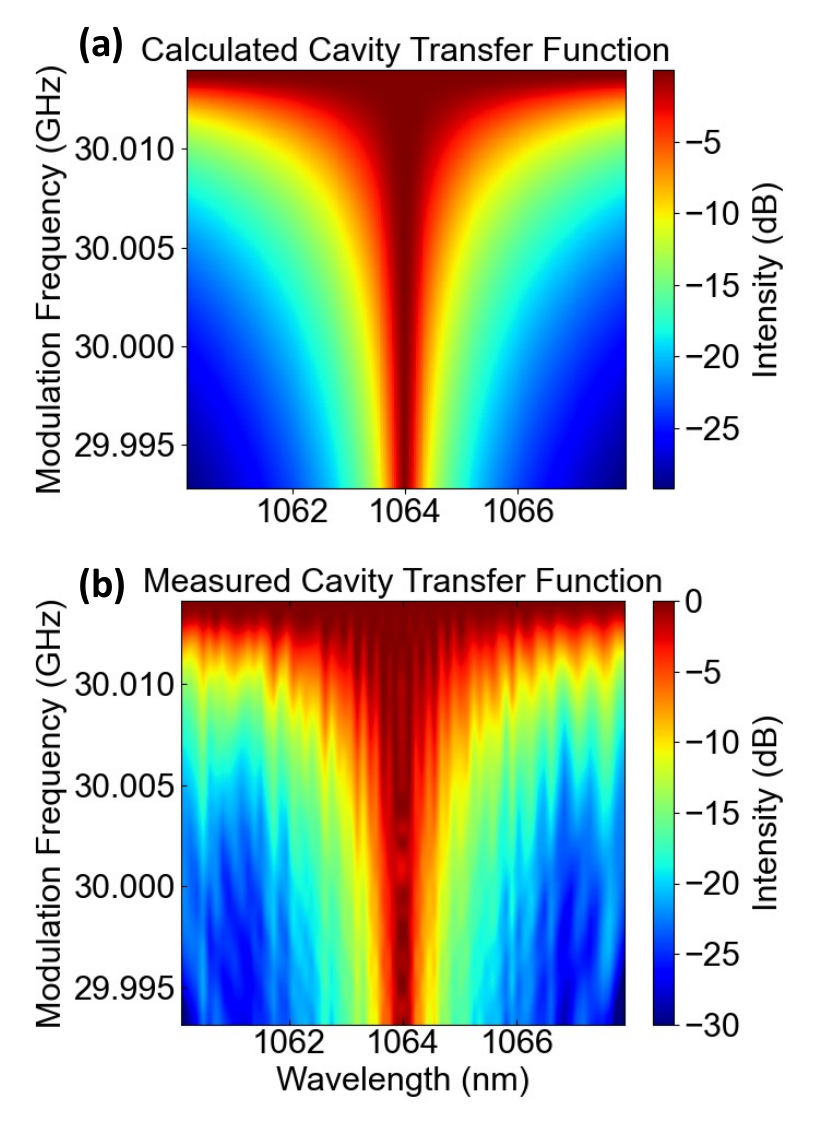}
\caption{Transmission of EO comb through the cavity on detuning the comb repetition rate from the cavity FSR - (a) calculated, and (b) measured. The calculated model shows good agreement with the experiment.}
\label{fig:4}
\end{figure}

Fig. \ref{fig:3} illustrates the technique used to achieve continuous comb tuning by 30 GHz. As the laser frequency is tuned, the cavity lock re-adjusts the cavity length to maximize the transmission of the center comb tooth. This results in a comb-cavity mode walk-off (see (iii) in Fig. \ref{fig:3}) since the FSR is no longer equal to the comb repetition rate. To avoid this walk-off, the comb repetition rate needs to be simultaneously tuned to match the new cavity FSR. This tandem tuning is well-described by a simple model of the transmission of the EO comb through the cavity given by, 
\begin{equation}
T=\frac{\left(1-R\right)^{2}}{1+R^{2}-2 R \cos \left(\frac{2 \omega l}{c}+\phi\right)}  
\label{eqn1}
\end{equation}
\begin{figure*}[htb]
\centering\includegraphics[width=0.9\linewidth]{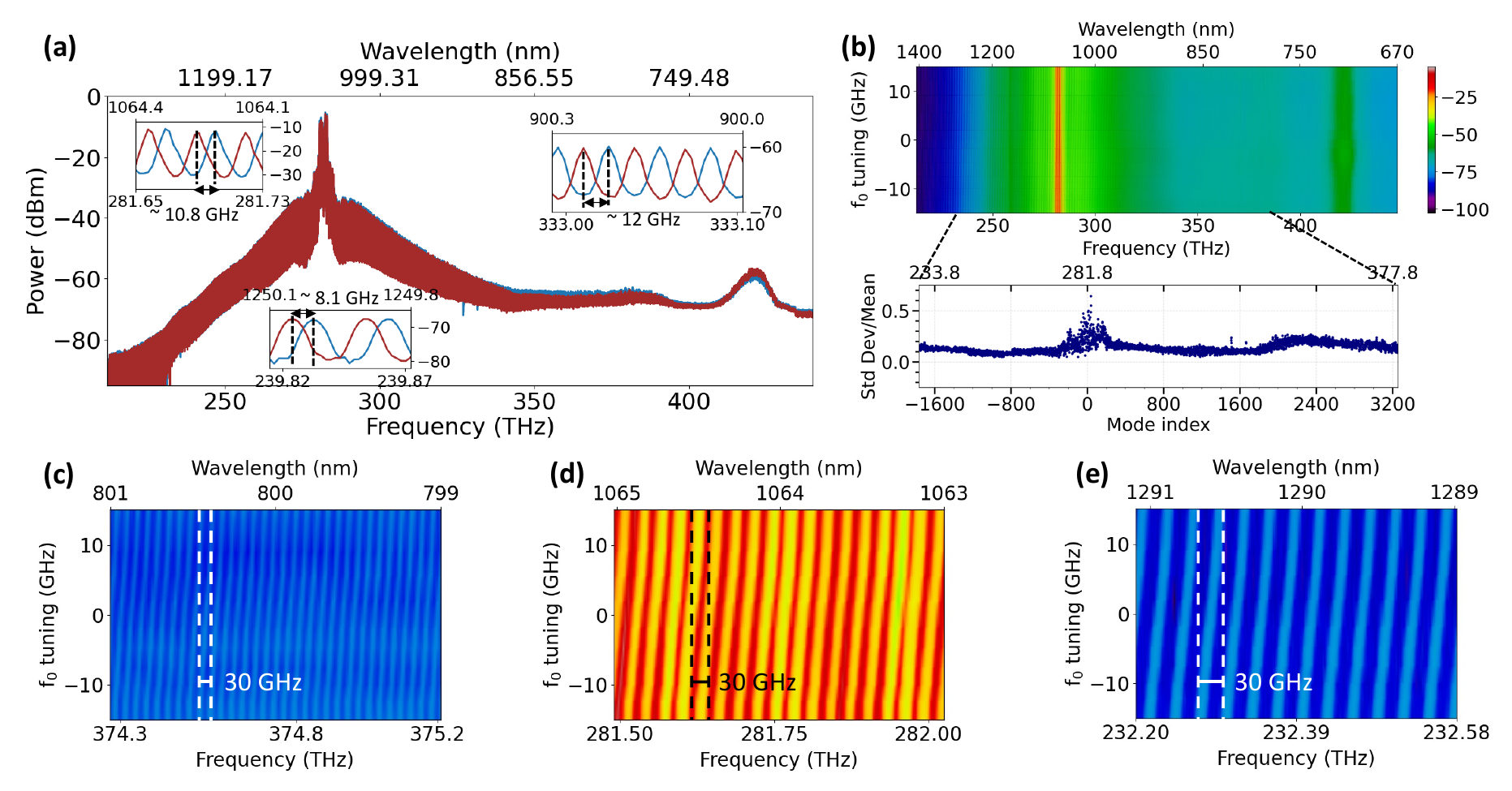}
\caption{Tuning 30 GHz supercontinuum spanning 700 - 1350 nm in SiN waveguide. (a) Exemplary comb-resolved supercontinua (SC) obtained by tuning the CW laser frequency through 10.8 GHz. (b) Coherent SC in logarithmic scale across the 30 GHz tuning range. Note here the SC remains mostly unchanged. The bottom plot shows the fractional intensity fluctuations in linear scale during the comb frequency tuning process. Zoomed-in version of (b) showing the linear frequency control of individual comb lines centered at (a) 800 nm, (b) 1064 nm, and (c) 1290 nm.}
\label{fig:6}
\end{figure*}
Here $R$ is the reflectivity of the cavity mirrors, $l$ is the length of the cavity, and $\phi$ is the phase-shift from the mirrors \cite{lawrence1999dynamic}. To achieve the full tunability of 30 GHz, the comb repetition rate needs to be changed by approximately 3.2 MHz.  Fig. \ref{fig:4} compares the model of the cavity transfer function from equation (\ref{eqn1}) with the measured one upon detuning the comb repetition rate from the cavity FSR. The good agreement between the calculated and measured cavity transmission further validates the technique in Fig. \ref{fig:3}. We also numerically investigated the effect of mirror dispersion on comb-cavity mode walk-off and found it to be negligible for a 10 nm bandwidth. The maximum change in cavity FSR, assuming dielectric mirrors comprised of a quarter-wave stack, is about 70 kHz, which is six orders of magnitude smaller than 30 GHz. 

The dispersion-compensated output of the cavity is amplified to 4.4 W using a ytterbium-doped fiber amplifier (YDFA) and launched into a temporal compressor consisting of an 8.5 m length of polarization-maintaining normal dispersion photonic crystal fiber (PCF) and grating compressor to yield a pulse duration of 113 fs (Fig. \ref{fig:2} (c)).
The compressed pulse is then launched into a dispersion-engineered silicon nitride (SiN) waveguide for the second stage of supercontinuum generation. The waveguide fabricated by the open-access foundry Ligentec is silica-clad, 25 mm long with a core cross-section of 690 nm × 800 nm. The calculated anomalous dispersion is 73 ps/nm/km at 1.064 $\mu$m. The total insertion loss is measured to be approximately 6.3 dB with an average power of 1.7 W incident on the waveguide. 
Fig. \ref{fig:6}(a) shows the smooth supercontinuum (SC) spanning 700 - 1300 nm generated in the SiN waveguide with 13 pJ pulse energy at two different instances when CW laser frequency is tuned by 10.8 GHz. The tuning technique is performed as described before. During the tuning process, the group-velocity dispersion parameter in the pulse shaper is finely adjusted and the input coupling to the waveguide at Watt level powers is maintained by piezo servos to retain the same envelope for SC measured on an optical spectrum analyzer (OSA). As a result, the broadband coherent SC generated in the nanophotonic SiN waveguide with a dispersive wave around 700 nm remains mostly unchanged across the entire 30 GHz tuning range. The fractional amplitude fluctuations on the SC across the 30 GHz tuning range when the CW laser frequency is tuned in steps of 3 GHz are shown in Fig. \ref{fig:6}(b). The maximum intensity fluctuations of 50$\%$ around the pump wavelength are attributed to our use of a free-running CW laser. The short-term intensity stability outside this region is about 10 - 20$\%$. The zoomed-in high-resolution plots in Figs. \ref{fig:6}(c), (d) and (e) show the linear change in frequency of the 30 GHz comb modes at center wavelengths of 800 nm, 1064 nm, and 1290 nm while tuning.  

To the best of our knowledge, this is the first demonstration of full FSR spectral tuning with a broadband comb having mode spacing appropriate for astronomical spectrograph calibration.  In doing so, we overcome the technical challenges of maintaining microwave noise reduction with a filter cavity, retaining the short $\sim$ 120 fs pulse, and generating broadband supercontinuum with constant spectral density. The same techniques can be implemented with other astrocomb architectures, and the full tunability will enable precise characterization of large format detector arrays as a critical step towards cm/s RV precision in NIR and visible. 

\begin{backmatter}
\bmsection{Funding} National Science Foundation (AST 2009889, AST 2009982, AST 2009955, AST 2009554).

\bmsection{Acknowledgments} 
We thank Michael Geiselmann of Ligentec for providing the SiN waveguides. The Center for Exoplanets and Habitable Worlds is supported by the Pennsylvania State University and the Eberly College of Science.

\bmsection{Disclosures} 
The authors declare no conflicts of interest.

\bmsection{Data Availability Statement} 
Data underlying the results presented in this paper may be obtained from the authors upon reasonable request.


\end{backmatter}




\bibliography{ref}

\bibliographyfullrefs{ref}

\end{document}